\documentclass[twocolumn,english]{revtex4-1}
\usepackage[T1]{fontenc}
\usepackage[latin9]{inputenc}
\usepackage{color}
\usepackage{textcomp}
\usepackage{mathrsfs}
\usepackage{amsmath}
\usepackage{amssymb}
\usepackage{graphicx}

\makeatletter

\newcommand{\lyxmathsym}[1]{\ifmmode\begingroup\def\b@ld{bold}
  \text{\ifx\math@version\b@ld\bfseries\fi#1}\endgroup\else#1\fi}

\providecommand{\tabularnewline}{\\}

\makeatother

\usepackage{babel}
\begin{document}
\title{Simultaneous Perfect Bending and Polarization Rotation of Electromagnetic
Wavefront using Chiral Gradient Metasurfaces}
\author{Hamidreza Kazemi}
\email{hkazemiv@uci.edu}

\author{Mohammad Albooyeh}
\email{mohammad.albooyeh@gmail.com}

\author{Filippo Capolino}
\email{f.capolino@uci.edu}

\affiliation{Department of Electrical Engineering and Computer Science, University
of California, Irvine, California 92697, United States}
\begin{abstract}
We introduce chiral gradient metasurfaces that allow perfect transmission
of all the incident wave into a desired direction and simultaneous
perfect rotation of the polarization of the refracted wave with respect
to the incident one. Besides using gradient polarization densities
which provide \textit{bending} of the refracted wave with respect
to the incident one, using metasurface inclusions that are \textit{chiral}
allows the \textit{polarization} of the refracted wave to be rotated.
We suggest a possible realization of the proposed device by discretizing
the required equivalent surface polarization densities realized by
proper helical inclusions at each discretization point. By only using
a single optically thin layer of chiral inclusions, we are able to
unprecedentedly deflect a normal incident plane wave to a refracted
plane wave at $45\lyxmathsym{\textdegree}$ with $72\%$ power efficiency
which is accompanied by a $90^{\circ}$ polarization rotation. The
proposed concepts and design method may find practical applications
in polarization rotation devices at microwaves as well as in optics,
especially when the incident power is required to be deflected.
\end{abstract}
\maketitle

\section{INTRODUCTION}

From the beginning of the 21st century, the investigation of metasurfaces,
i.e., optically thin layers of arrayed subwavelength inclusions, to
shape the wavefront of the electromagnetic waves at will is dramatically
increased compared to that of bulky metamaterials \cite{albooyeh_electromagnetic_2016,achouri_general_2015,asadchy_perfect_2016,chen_review_2016,cheng_ultrathin_2013,faniayeu_highly_2017,markovich_metamaterial_2013,menzel_asymmetric_2010,pfeiffer_bianisotropic_2014,pfeiffer_polarization_2016,safari_cylindrical_2017,vehmas_eliminating_2013,xiao_multi-band_2015,zheng_wideband_2018,zhu_manipulating_2017,Capolino2013,Glybovski2016,Tretyakov2015,Wong2016,Kazemi2019chiral,MohammadiEstakhri2016,Alaee:16,Albooyeh_Revisting-2015,guo_topologically_2017,rajaei2019giant}.
This is because metasurfaces have in general less losses and easier
manufacturing processes compared to bulky engineered metamaterials.
Quite recently, by applying the so-called generalized laws of reflection
and refraction, specifically designed phase gradient metasurfaces
achieved about $25\%$ of transmitted power for manipulation of transmitted
waves \cite{Yu2011,Shalaev2015}. Such designs were suffering from
a lack of degrees of freedom for controlling the polarization of the
refracted wave. Subsequent attempts, based on generalized boundary
conditions, accomplished more efficient power operation (about $80\%$)
and also enabled manipulation of polarization \cite{Monticone2013,Pfeiffer2013}.
Simultaneous control of reflected or transmitted phase and amplitude
is achieved using anisotropic metasurface elements with enough degrees
of freedom as the Y shaped elements in \cite{veysi_thin_2015} leading
to both deflection and polarization rotation, yet without devising
a robust method that maximizes power transfer.

Most recently, a theoretical scheme for gradient (spatially dispersive)
metasurfaces, which offers a perfect control of refracted/reflected
waves (i.e., 100\% power efficiency), was introduced in the seminal
studies in Refs. \cite{asadchy_perfect_2016,MohammadiEstakhri2016,Wong2016}.
Based on that scheme, metasurface designs several interesting applications
involving wavefront control were proposed (see e.g., Refs. \cite{Elsakka2016,Chen2017,Asadchy2017,Emani2017,Wong2018,rabinovich2018analytical}).
However, yet no work has been carried out based on that scheme for
the concurrent control of both the direction and polarization of refracted
wavefront. Therefore, we found it timely and essential to introduce
the realization of gradient metasurfaces that not only desirably deflect
the wavefront but also manipulate its polarization at will. 

Metasurfaces for both polarization conversion and wavefront manipulation
may find a wide range of applications in problems in frequency ranges
spanning from microwaves to optics. For instance, metasurface based
polarization rotators are suitable to replace bulky wave plates (quarter,
half wavelength, etc.). Next, metasurface based wave deflectors are
handy candidates to take over the commonly used bulky optical beam
splitters that deflect the wavefront of light ($45\lyxmathsym{\textdegree}$deflection
with $50\%$ power efficiency) in optical systems \cite{pfeiffer_bianisotropic_2014,yu_broadband_2012,wu2014spectrally}.
Moreover, polarization selective metasurfaces are applied for coding
the information into different polarization states \cite{Selvanayagam2016,li_optical_nodate,kim2016highly}.
Furthermore, they may find practical applications as both polarization
rotators or wavefront deflectors at microwave frequencies especially
in the design of antenna systems, etc \cite{Selvanayagam2016,pfeiffer_polarization_2016,jia_broadband_2016,farmahini-farahani_birefringent_2013}.

Here, we synthesize a planar transmitting metasurface which perfectly
deflects the normal (with respect to the metasurface plane) incoming
wavefront by $45\lyxmathsym{\textdegree}$ and concurrently rotates
its polarization by $90\lyxmathsym{\textdegree}$, while we express
that a similar design procedure can be performed for any arbitrary
angles of deflection and any polarization rotation. Our synthesis
approach is based on the realization of the desired electric and magnetic
equivalent surface polarization densities which are connected to the
total fields at both sides of the metasurface through the ``sheet
boundary conditions'' as described in Sec. \ref{sec:PROBLEM-DESCRIPTION}.
We further analyze the metasurface performance for the obtained electric
and magnetic polarization densities when they are discretized into
five sampling points for each supercell. In Sec. \ref{sec:ExamplePHYSICAL-REALIZATION-EXAM}
we present a physical realization design that almost satisfies the
required equivalent polarization densities which were obtained in
the previous section. The design is composed of five unit cells which
form a supercell. Each unit cell consists of four interlaced helices
which fulfills the desired polarization densities, when judiciously
engineered. The performance of this metasurface design is compared
with the ideal case, and a power transmission efficiency of $72\%$
is achieved for a $90\lyxmathsym{\textdegree}$ polarization rotation
and a $45\lyxmathsym{\textdegree}$ wavefront deflection which is
much higher than $25\%$ efficiency reported in the literature for
a smaller angle of wavefront deflection (e.g., $30\lyxmathsym{\textdegree}$)
using the so-called generalized laws of reflection and refraction
(see e.g., Ref. \cite{aieta_out--plane_2012}).

We emphasize that besides being electromagnetically thin, our proposed
strategy offers two simultaneous functionalities, i.e., wavefront
bending along with polarization rotation of the incoming wavefront,
in a single layer design with an unprecedentedly high efficiency.
Such a device dramatically reduces the required space by combining
the advantages of these two functionalities. Indeed, it can be used
in place of two most commonly-used optical apparatus i.e., a wave
plate and a beam splitter which are bulky and occupy substantial spaces.
Moreover, our proposal has the advantage of perfect performance (refraction
and polarization\textcolor{black}{{} rotation) when compared to these
devices.}

\section{PROBLEM DESCRIPTION\label{sec:PROBLEM-DESCRIPTION}}

\begin{figure}[t]
\begin{raggedright}
(a)~~~~~~~~~~~~~~~~~~~~~~~~~~~~~~~~~~~~~~~~~~~~~(b)
\par\end{raggedright}
\begin{centering}
\includegraphics[width=1.75in]{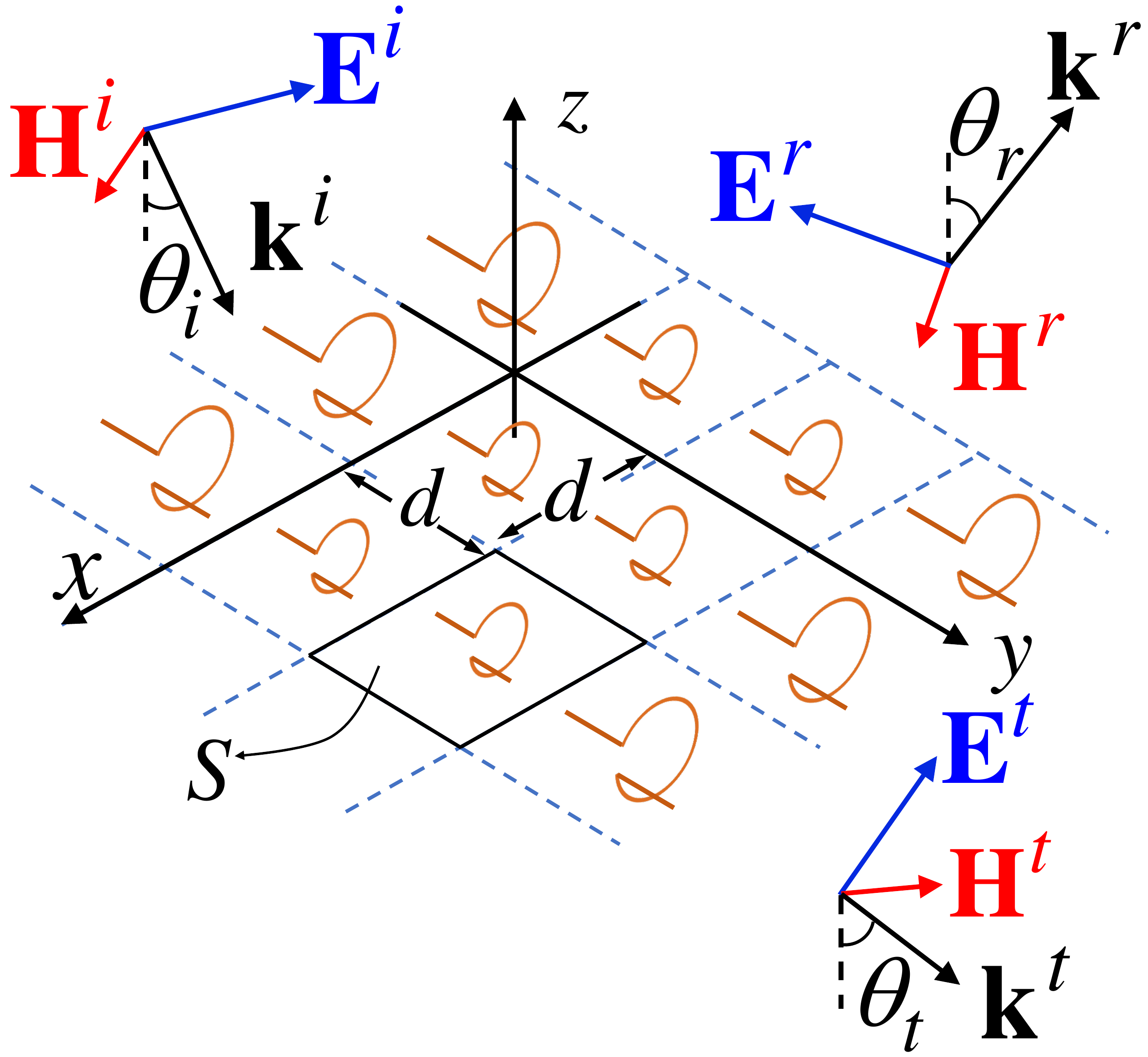}\includegraphics[width=1.75in]{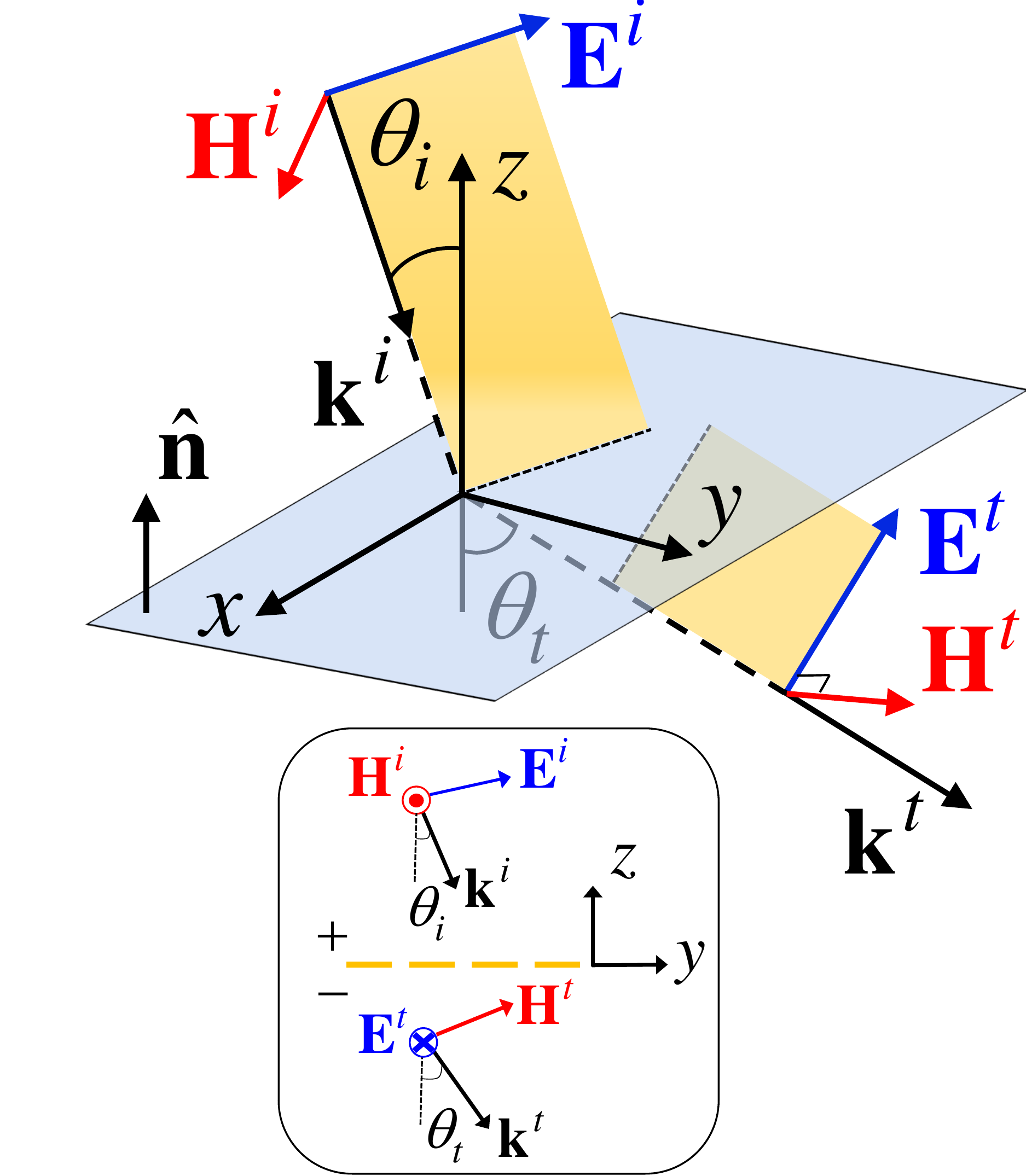}\centering
\par\end{centering}
\centering{}\caption{(a) The general gradient metasurface and depiction of incident, reflected
and transmitted waves. (b)The incident and refracted waves and polarization
of the fields for a reflectionless metasurface.}
\label{Chiral_MS}
\end{figure}

Let us consider a general metasurface that \textit{perfectly} refracts
an incident plane wave with incoming angle $\theta_{i}$ to a plane
wave with a desired refraction angle $\theta_{t}$, and converts the
incident field polarization to a favorable one in the refracted field
{[}see Fig. \ref{Chiral_MS}{]}. 
To elaborate the concepts in a simple manner, let us now make our
analysis more specific and consider an incident plane wave with a
transverse magnetic (TM with respect to $z$) polarization which is
going to be refracted as a transverse electric (TE) polarized plane
wave as shown in the subset of the Fig. \ref{Chiral_MS}(b). In this
example the polarization of the refracted wave is hence rotated by
$90\lyxmathsym{\textdegree}$ with respect to that of the incident
wave. Next, assuming time a harmonic wave with time dependence $e^{j\omega t}$
which travels downward, i.e., in the $-z$ direction {[}see Fig. \ref{Chiral_MS}(b){]},
the electric field vector of the incident wave reads $\mathbf{E}^{i}=E_{0}^{i}\left(\hat{{\bf y}}\cos\theta_{i}{\rm e}^{-jk_{0}(\sin\theta_{i}y-cos\theta_{i}z)}\right.$
$\left.+\hat{{\bf z}}\sin\theta_{i}{\rm e}^{-jk_{0}(sin\theta_{i}y-\cos\theta_{i}z)}\right)$
and that of the refracted wave reads $\mathbf{E}^{t}=-\hat{{\bf x}}E_{0}^{t}{\rm e}^{-jk_{0}(\sin\theta_{t}y-cos\theta_{t}z)-j\varphi_{t}}$.
Here, $\hat{{\bf x}}$, $\hat{{\bf y}}$, and $\hat{{\bf z}}$ are
the unit vectors in Cartesian coordinates, whereas $y$ and $z$ are
accordingly the position variables. Moreover, $k_{0}$ is the free-space
wave number, $E_{0}^{i}$ and $E_{0}^{t}$ are the electric field
amplitudes of the incident and refracted waves, respectively, and
$\varphi_{t}$ accounts for a possible phase shift between the incident
and refracted waves upon crossing the metasurface. The tangential
electric $\mathbf{E}_{t}$ and magnetic $\mathbf{H}_{t}$ fields (the
subscript $t$ denotes the tangential component with respect to the
metasurface plane) at the boundary of the metasurface, on both the
upper and the lower sides, assuming the metasurface is located on
the $z=0$ plane, read 

\begin{align}
\mathbf{E}_{t+} & =\hat{{\bf y}}E_{0}^{i}\cos\theta_{i}e^{-jk_{0}\sin\theta_{i}y},\label{E_t}\\
\mathbf{E}_{t-} & =-\hat{{\bf x}}E_{0}^{t}e^{-jk_{0}\sin\theta_{t}y-j\varphi_{t}},\nonumber 
\end{align}

\begin{align}
\hat{\mathbf{n}}\times\mathbf{H}_{t+} & =\hat{{\bf y}}\frac{E_{0}^{i}}{\eta_{0}}e^{-jk_{0}\sin\theta_{i}y},\label{H_t}\\
\nonumber \\
\hat{\mathbf{n}}\times\mathbf{H}_{t-} & =-\hat{{\bf x}}\frac{\cos\theta_{t}}{\eta_{0}}E_{0}^{t}e^{-jk_{0}\sin\theta_{t}y-j\varphi_{t}},\nonumber 
\end{align}
respectively. Here, ``$t+$'' and ``$t-$'' subscripts refer to
the tangential fields at the upper ($z>0$) and lower ($z<0$) metasurface
boundaries, respectively. Furthermore $\eta_{0}=\sqrt{\mu_{0}\epsilon_{0}}$
is the intrinsic wave impedance of the free space, and $\hat{\mathbf{n}}$
is the unit vector normal to the metasurface plane in the $+z$ direction
{[}see Fig.\ref{Chiral_MS}(b){]}. From Eq. (\ref{E_t}), the phase
of the transmission coefficient reads

\begin{eqnarray}
\Phi_{t} & = & k_{0}\left(\sin\theta_{i}-\sin\theta_{t}\right)y+\varphi_{t},\label{trans_phase}
\end{eqnarray}
which is obviously not uniform over the surface since $\theta_{i}\neq\theta_{t}$
and indeed it varies linearly with $y$. (Other nonlinear variation
may be required for other applications, e.g., focusing \cite{khorasaninejad2015achromatic,veysi_thin_2015}.)

In the next step, we write the boundary conditions which connect the
jump of tangential fields on both sides of the metasurface to the
induced equivalent electric $\mathbf{P}$ and magnetic $\mathbf{M}$
surface polarization densities as \cite{albooyeh_electromagnetic_2016,albooyeh_normal_2017,albooyeh_equivalent_2017}

\begin{eqnarray}
\mathbf{E}_{t+}-\mathbf{E}_{t-} & = & j\omega\hat{\mathbf{n}}\times\mathbf{M},\label{E_BC}\\
\hat{\mathbf{n}}\times\mathbf{H}_{t+}-\hat{\mathbf{n}}\times\mathbf{H}_{t-} & = & j\omega\mathbf{P},\label{H_BC}
\end{eqnarray}
where $\omega$ is the angular frequency. Ther\textcolor{black}{efore,
}by plugging Eqs. (\ref{E_t}), (\ref{H_t}), and (\textcolor{black}{\ref{trans_phase}})
into Eqs. (\ref{E_BC}) and (\ref{H_BC}), the required electric and
magnetic equivalent surface polarization densities for the proposed
field manipulation, respectively, read 

\begin{eqnarray}
\mathbf{P} & = & \frac{E_{0}^{i}}{j\omega\eta_{0}}{\rm e}^{-jk_{0}\sin\theta_{i}y}\left[-\cos\theta_{t}\,t_{xy}e^{-j\Phi_{t}}\hat{{\bf x}}+\hat{{\bf y}}\right],\label{pt_general}\\
\nonumber \\
\mathbf{M} & = & \frac{E_{0}^{i}}{j\omega}{\rm e}^{-jk_{0}\sin\theta_{i}y}\left[\cos\theta_{i}\hat{{\bf x}}-t_{xy}e^{-j\Phi_{t}}\hat{{\bf y}}\right].\label{mt_general}\\
\nonumber 
\end{eqnarray}

Here, the TM to TE polarized wave transmission coefficient $t_{xy}=E_{x}^{t}/E_{y}^{i}=E_{0}^{t}/\left(E_{0}^{i}\cos\theta_{i}\right)$,
and by neglecting losses it is approximated as (see Appendix \ref{AppB}
for a proof)

\begin{equation}
t_{xy}=\frac{1}{\sqrt{\cos\theta_{i}\cos\theta_{t}}},
\end{equation}
where we consider $E_{x}^{t}=E_{0}^{t}$ and $E_{y}^{i}=E_{0}^{i}\cos\theta_{i}$.
Note that the transmission coefficient here is defined with respect
to the transverse component of the electric field in the two half
spaces, thus, $t_{xy}$ can be larger than unity for an oblique incident
or transmission angle, without contradicting the power conservation
law. 

\subsection{Illustrative example}

Let us now consider a specific example where $\theta_{i}=0$ and $\theta_{t}=\pi/4$
and with each unit cell having the dimension $d$ along $x$ and $D$
along $y$ axes as shown in Fig. \ref{desired_p,m}(a). Figure \ref{desired_p,m}(b)
and (c) show the required electric and magnetic equivalent polarization
densities described by Eqs. (\ref{pt_general}) and (\ref{mt_general})\textcolor{green}{{}
}\textcolor{black}{in a unit cell (supercell) for such a metasurface
with the mentioned characteristic, i.e., }$\theta_{i}=0$ and $\theta_{t}=\pi/4$\textcolor{black}{.}

\begin{figure}[t]
\begin{raggedright}
(a)
\par\end{raggedright}
\begin{centering}
\includegraphics[width=3.5in]{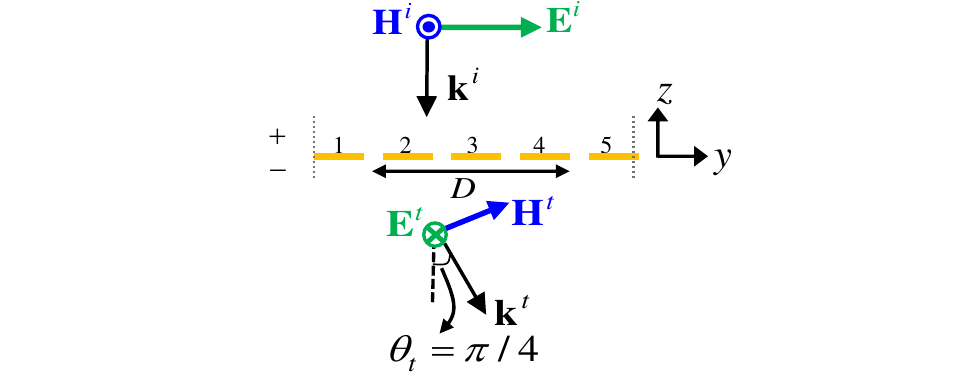}
\par\end{centering}
\begin{raggedright}
(b)~~~~~~~~~~~~~~~~~~~~~~~~~~~~~~~~~~~~~~(c)
\par\end{raggedright}
\begin{centering}
\includegraphics[width=3.5in]{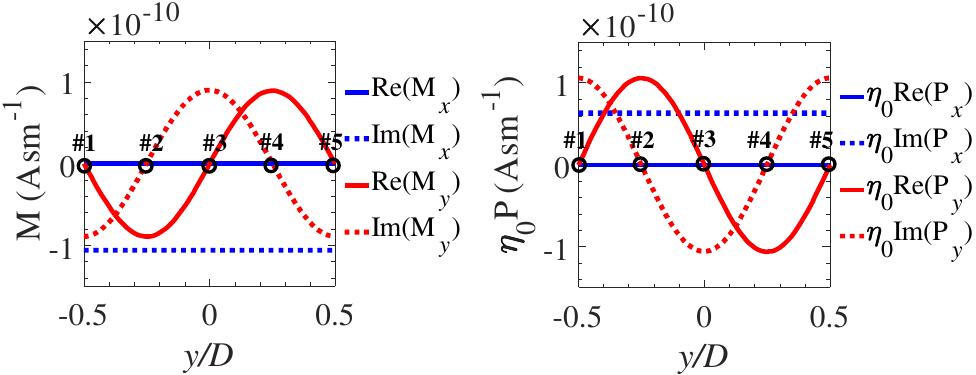}
\par\end{centering}
\centering{}\caption{(a) Schematic of the incident electric field polarized along $y$
($E_{y}$) normally propagating and of the transmitted wave propagating
at an angle from the normal direction with a rotated polarization
along $x$ ($E_{x}$). (b) The desired magnetic and (c) electric polarization
densities of a supercell for $y$-polarized to $x$-polarized electric
field conversion in transmission. The black circles in this figure
show the discretization points of the polarization densities on the
$y$ axis.}
\label{desired_p,m}
\end{figure}
 The obtained equivalent polarization surface densities are continuous,
uniform in the $x$ direction, and periodic in the $y$ direction
(with period $D$). However, in general a metasurface is practically
composed of discrete elements that mimic such continuous polarization
densities. In order to realize a practical design with discrete elements,
we take a finite number of sampling points (here five equi-distance
points, $D=5d$) in $y$ direction within each supercell to sample
the desired continuous equivalent polarization densities given in
Fig. \ref{desired_p,m}(b), (c). To demonstrate how the discretization
procedure affects the performance of the design we substitute the
equivalent polarization densities with electric and magnetic point
sources with equivalent magnitudes and phases as sampled at these
five sampling points. By implementing the electric and magnetic point
sources which mimic the required continuous equivalent polarization
densities, we plot the simulation results (using the finite element
method implemented in COMSOL multiphysics \cite{COMSOL}) of the electric
field distribution: the $y$-component of the incident electric field
and the $x$-component of the refracted field, in Fig. \ref{fig:(a)-Geometry,-incident,}. 

\begin{figure}[t]
\begin{centering}
\includegraphics[width=3.45in]{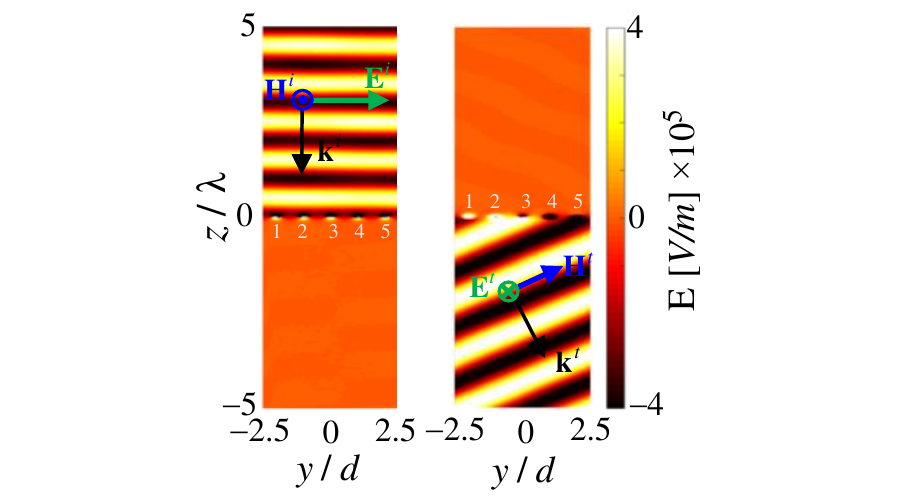}
\par\end{centering}
\centering{}\caption{\label{fig:(a)-Geometry,-incident,} Electric field distribution (snapshot
at a given time) of a perfectly refracting and polarization rotating
metasurface on both sides of the metasurface. The result is based
on calculating the scattered field as generated by electric and magnetic
point-dipole sources evaluated of five sampling points. The incident
field is polarized along $y$ ($E_{y}$) and does not experience reflection,
i.e., no standing wave is present. The transmitted electric field
is polarized along $x$ ($E_{x}$) and it is refracted at an oblique
angle $\theta_{t}=45^{\circ}$. Total transmission is achieved, i.e.,
there is no reflection.}
\end{figure}
It is clear from the field distribution maps that the power is perfectly
refracted by $45^{\circ}$ and rotated by $90^{\circ}$, and moreover,
as expected the energy density of the refracted wave is lower than
the incident one \textcolor{black}{since the total power of the refracted
wave is imposed to be equal to the incident power }(see the Appendix
\ref{AppB} for more discussion about energy and power densities).
There is no reflection as it can be observed by the lack of any standing
wave.

\subsection{Possible physical topologies}

The result of the previous section indicates the required polarization
densities without any discussion about a feasible realization. In
order to understand possible topologies for the realization of a metasurface
that acquires the proposed equivalent polarization densities, it is
helpful to understand what kinds of scatterers (elements or inclusions)
are able to create such polarization densities. Considering inclusion
realizations that are well approximated by dipole moments, the electric
and magnetic polarization densities so far described are realized
by using electric $\mathbf{p}_{t}=\mathbf{P}S$ and magnetic $\mathbf{m}_{t}=\mathbf{M}S$
equivalent dipole polarizations per unit cell where $S$ is the unit
cell area (see Figure \ref{Chiral_MS}(a)). One practical tool is
to study the relation between the fields and the equivalent dipole
polarizations per unit cell, i.e., the constitutive relations. Indeed,
the tangential components of the equivalent dipole polarization densities
${\bf p}_{t}$ and ${\bf m}_{t}$ and the tangential incident field
components ${\bf E}_{t}^{i}$ and ${\bf H}_{t}^{i}$ are related through
tangential components of the collective polarizabilities $\bar{\bar{\boldsymbol{\alpha}}}_{t}^{\mathrm{ee}}$,
$\bar{\bar{\boldsymbol{\alpha}}}_{t}^{\mathrm{em}}$, $\bar{\bar{\boldsymbol{\alpha}}}_{t}^{\mathrm{me}}$,
and $\bar{\bar{\boldsymbol{\alpha}}}_{t}^{\mathrm{mm}}$ of the metasurface
unit cells via constitutive relations \cite{albooyeh_electromagnetic_2016,Serdiukov2001electromagnetics,hanifeh_helicity_2019}

\begin{eqnarray}
 & \mathbf{p}_{t}=\bar{\bar{\boldsymbol{\alpha}}}_{t}^{\mathrm{ee}}\mathbf{\cdot E}_{t}^{i}+\bar{\bar{\boldsymbol{\alpha}}}_{t}^{\mathrm{em}}\cdot\mathbf{H}_{t}^{i},\label{eq:p_constitutive}\\
 & \mathbf{m}_{t}=\bar{\bar{\boldsymbol{\alpha}}}_{t}^{\mathrm{me}}\mathbf{\cdot E}_{t}^{i}+\bar{\bar{\boldsymbol{\alpha}}}_{t}^{\mathrm{mm}}\cdot\mathbf{H}_{t}^{i}.\label{eq:m_constitutive}
\end{eqnarray}

In the above equations, $\bar{\bar{\boldsymbol{\alpha}}}_{t}^{\mathrm{ee}}$,
$\bar{\bar{\boldsymbol{\alpha}}}_{t}^{\mathrm{em}}$, $\bar{\bar{\boldsymbol{\alpha}}}_{t}^{\mathrm{me}}$,
and $\bar{\bar{\boldsymbol{\alpha}}}_{t}^{\mathrm{mm}}$ are collective
electric, magnetoelectric, electromagnetic, and magnetic polarizability
dyadics in the metasurface plane, relating the incident field to the
electric and magnetic dipoles. The word \textit{collective} means
that the polarizability accounts also for the coupling with all the
dipoles in the other unit cells of the array \cite{albooyeh_electromagnetic_2016,Serdiukov2001electromagnetics,albooyeh2018applications}.
Each polarization dyad has four components $\alpha_{ij}$, where $ij=xx,xy,yx,\,{\rm or\,}yy$
in Cartesian coordinates, i.e., in terms of matrix representation
is given by

\begin{equation}
\bar{\bar{\alpha}}_{t}=\left[\begin{array}{cc}
\alpha_{xx} & \alpha_{xy}\\
\alpha_{yx} & \alpha_{yy}
\end{array}\right].
\end{equation}

Note that in the general case polarizability dyadics have nine components,
however, for our analysis it is enough to consider only tangential
components (see Ref. \cite{albooyeh_normal_2017,albooyeh_equivalent_2017}
for a more elaborated discussion). Next, by plugging the fields from
Eqs. (\ref{E_t}) and (\ref{H_t}) into Eqs. (\ref{eq:p_constitutive})
and (\ref{eq:m_constitutive}), the equivalent dipole polarizations
in term of polarizability components and the incident field read

\begin{eqnarray}
\mathbf{p}_{t} & = & \left[\hat{{\bf x}}\left(\mathrm{\alpha}_{xy}^{\mathrm{ee}}\cos\theta_{i}+\frac{\mathrm{\mathrm{\alpha}}_{xx}^{\mathrm{em}}}{\eta_{0}}\right)\right.\label{p_alphas}\\
 &  & \left.+\hat{{\bf y}}\left(\mathrm{\alpha}_{yy}^{\mathrm{ee}}\cos\theta_{i}+\frac{\mathrm{\alpha}_{yx}^{\mathrm{em}}}{\eta_{0}}\right)\right]E_{0}^{i}{\rm e}^{-jk_{0}\sin\theta_{i}y},\nonumber \\
\mathbf{m}_{t} & = & \left[\hat{{\bf x}}\left(\mathrm{\alpha}_{xy}^{\mathrm{me}}\cos\theta_{i}+\frac{\mathrm{\alpha}_{xx}^{\mathrm{mm}}}{\eta_{0}}\right)\right.\label{m_alphas}\\
 &  & \left.+\hat{{\bf y}}\left(\mathrm{\mathrm{\alpha}}_{yy}^{\mathrm{me}}\cos\theta_{i}+\frac{\mathrm{\alpha}_{yx}^{\mathrm{mm}}}{\eta_{0}}\right)\right]E_{0}^{i}{\rm e}^{-jk_{0}\sin\theta_{i}y}.\nonumber 
\end{eqnarray}

Based on the above equations, there is obviously not a unique scatterer's
topology to deliver the desired performance since the above are four
scalar complex-value equations with 8 complex-value unknown polarizability
components (four in each polarizability dyad, however, the selected
polarizations in this particular case imply that only eight dyad entrees
need to be determined). Nevertheless, among all possible solutions,
both electric and magnetic polarizations must be simultaneously nonzero,
which this limits the number of possible solutions. Moreover, in the
following we suppress the cross-components of the polarizabilities
(i.e, $\mathrm{\alpha}_{xy}^{\mathrm{ee}}=\mathrm{\alpha}_{yx}^{\mathrm{em}}=\mathrm{\alpha}_{xy}^{\mathrm{me}}=\mathrm{\alpha}_{yx}^{\mathrm{mm}}=0$)
to narrow down the possible sets of topologies. As a result, the remaining
polarizability components imply that the metasurface constitutive
inclusions must be chiral since we require that $\mathrm{\alpha}_{xx}^{\mathrm{em}}=\mathrm{-\alpha}_{xx}^{\mathrm{me}},\neq0$
or $\mathrm{\alpha}_{yy}^{\mathrm{em}}=\mathrm{-\alpha}_{yy}^{\mathrm{me}},\neq0$,
for reciprocal lossy and lossless inclusions. By using Eqs. (\ref{pt_general})
and (\ref{mt_general}) in Eqs. (\ref{p_alphas}) and (\ref{m_alphas})
the collective polarizabilities for the proposed chiral metasurface
are

\begin{eqnarray}
\mathrm{\alpha}_{xx}^{\mathrm{em}} & = & -\frac{S}{j\omega}\cos\theta_{t}t_{xy}e^{-j\Phi_{t}},\nonumber \\
\mathrm{\alpha}_{yy}^{\mathrm{ee}} & = & \frac{S}{j\omega}\frac{1}{\eta_{0}\cos\theta_{i}},\label{perfect_T_alphas}\\
\mathrm{\alpha}_{xx}^{\mathrm{mm}} & = & \frac{S}{j\omega}\eta_{0}\cos\theta_{i},\nonumber \\
\mathrm{\alpha}_{yy}^{\mathrm{me}} & = & -\frac{S}{j\omega}\frac{t_{xy}}{\cos\theta_{i}}e^{-j\Phi_{t}}.\nonumber 
\end{eqnarray}

In the next section we propose a physical element that exhibits these
polarizability components, i.e., the desired equivalent dipole polarizations,
and hence, the required equivalent surface polarization densities
under the given illumination.

\section{PHYSICAL REALIZATION\label{sec:ExamplePHYSICAL-REALIZATION-EXAM}}

In this section, we first propose a unit cell design that perfectly
rotates the polarization of the incoming wave at a normal incidence
by $90^{\circ}$. Hence, in the first step, we do not generate plane
wave deflection. This latter feature will be realized as a second
step by applying the designed unit cell of this first step and constructing
a gradient metasurface, to achieve a perfect deflection of a normal
incident wave to a $45^{\circ}$ refracted wave.

\subsection{Unit cell design for polarization rotation\label{subsec:Unit-cell-design}}

\begin{figure}[t]
\begin{raggedright}
(a)~~~~~~~~~~~~~~~~~~~~~~~~~~~~~~~~~~~~~(b)
\par\end{raggedright}
\begin{centering}
\includegraphics[width=1.5in]{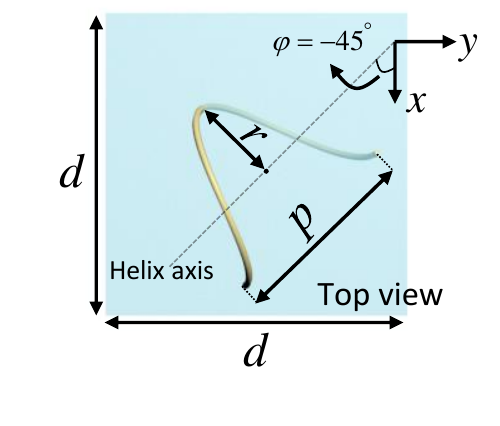}\includegraphics[width=1.5in]{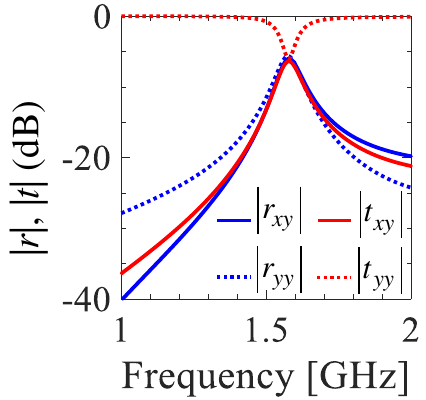}
\par\end{centering}
\begin{raggedright}
(c)~~~~~~~~~~~~~~~~~~~~~~~~~~~~~~~~~~~~~(d)
\par\end{raggedright}
\begin{centering}
\includegraphics[width=1.5in]{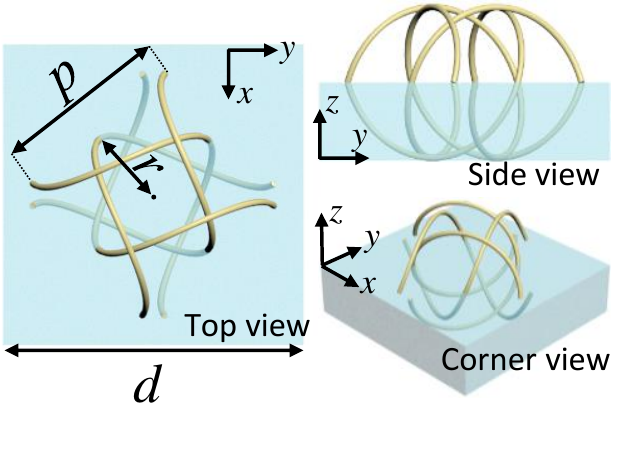}\includegraphics[width=1.5in]{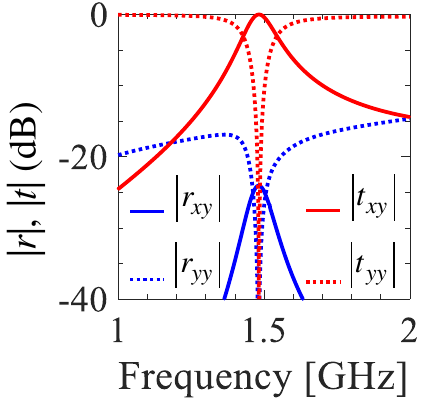}\centering
\par\end{centering}
\caption{\label{fig:(a)-Geometry-and}(a) The top view of a single-helix unit
cell with subwavelength dimensions. (b) Reflection and transmission
coefficients of\textcolor{red}{{} }$y$- and $x$-polarized waves for
the infinite planar periodic array of unit cells in (a). (c) Geometry
and different view angles of the designed particle composed by four
interlaced helices as in (a). (d) Reflection and transmission coefficients
of the\textcolor{red}{{} }$y$- and $x$-polarized waves for an infinite
planar periodic array of unit cells shown in (c).}
\end{figure}

It is well-known that a helical wire particle if excited properly
may acquire the polarizabilities described by Eqs. (\ref{perfect_T_alphas})
(see Ref. \cite{Tretyakov1996,fernandez2019new}). Here we use helices
with axes belonging to the transverse metasurface plane since we are
interested in inducing transverse electric and magnetic dipoles. To
optimize the performance of a chiral particle Semchenko et al. introduced
optimal helices in Ref. \cite{semchenko_optimal_2009} (i.e., helices
with equal electric, magnetic, and magneto-electric polarizabilities
which means $\left|\mathrm{\alpha}^{\mathrm{em}}\right|=\left|\eta_{0}\mathrm{\alpha}^{\mathrm{ee}}\right|=\left|\mathrm{\alpha}^{\mathrm{mm}}/\eta_{0}\right|$)
which we use here as a part of the design. Such an inclusion provides
a maximum cross-polarized transmission when implemented as the building
block of a metasurface. The basic building block of a metasurface
unit cell is a single optimal helix shown in Fig. \ref{fig:(a)-Geometry-and}(a).
In particular we consider a single helix made of one turn. The structural
parameters i.e., helices radius $r=14\,\mathrm{mm}$, helix pitch
$p=16.3\,\mathrm{mm}$, unit cell period $d=56.5\,\mathrm{mm}$ of
the designed helices are illustrated in Fig. \ref{fig:(a)-Geometry-and}(a).
Moreover the helix axis is oriented along the $\varphi=-45^{\circ}$
direction as shown in Fig. \ref{fig:(a)-Geometry-and}(a), and the
metal is assumed to be a perfect conductor with radius $r_{w}=0.1\,\mathrm{mm}$.

The co-polarized (i.e., along the same direction as the incident electric
field) and cross-polarized (i.e., orthogonal to the direction of the
incident electric field) field components that defined the reflection
and transmission coefficients of a metasurface composed of the unit
cell with single helix in Fig. \ref{fig:(a)-Geometry-and}(a) are
shown in Fig. \ref{fig:(a)-Geometry-and}(b). The subscripts $yy$
and $xy$ refer to co and cross polarization components, respectively,
in a linear polarization basis with the incident electric polarization
oriented along the $y$ direction. As stated earlier, in the first
step we seek a metasurface that is able to perfectly refract the normal
incident plane with $90^{\circ}$ rotation of the polarization with
respect to that of the incident wave, without generating any angular
deflection, i.e., the transmitted wave is propagating along the normal
direction (the\textit{ $z$} direction). In terms of reflection/transmission
coefficients, it means that
\begin{eqnarray}
t_{xy}= & 1\label{eq:refless_MS}\\
t_{yy}= & r_{xy}= & r_{yy}=0,\nonumber 
\end{eqnarray}
 As it is clear from Fig. \ref{fig:(a)-Geometry-and}(b), such a design
although provides a maximum possible cross-polarized transmission
with individual single-turn helix as unit cell, it does not grant
perfect transmission of all the incident power to the desired rotated
polarization (see that $|t_{xy}|\neq1$). Indeed, the incident power
is approximately shared evenly between all components of the reflection
and transmission spectra at the desired frequency (here around $1.5\,\mathrm{GHz}$),
hence Eqs. (\ref{eq:refless_MS}) are not satisfied. However, when
we increase the number of helices in a unit cell with a proper orientation
as mentioned in Ref. \cite{Asadchy2015}, i.e., four helices which
are rotated by 90 degrees around the $z$ axis with respect to each
other, then, remarkably, a reflectionless surface can be achieved.
Nevertheless, the design in Ref. \cite{Asadchy2015} has a unit cell
size that exceeds the operational wavelength, hence it is not a practical
design for metasurfaces with incident (refracted) angles rather than
normal. Therefore, here we use four interlaced helices in a single
unit cell to be sure the unit cell size is subwavelength, a feature
that is useful for the implementation of a gradient metasurface which
generates a transmitted wave with a $45^{\circ}$ deflection, discussed
in the next subsection. Figure \ref{fig:(a)-Geometry-and}(c) shows
the configuration of the four interlaced helices in each unit cell
that provide a fully transmittive (i.e., reflectionless) metasurface
which perfectly rotates the polarization of the normal incoming wave
by $90^{\circ}$. This unit cell is composed of four identical co-centered
helices in which their axes lie on the $xy$-plane and they are rotated
by 90 degrees around the $z$ axis. The orientation, spatial position,
and structural parameters of the designed helices are illustrated
in Figure \ref{fig:(a)-Geometry-and}(c), where the structural parameters
for each helix are the same as those for the single-helix unit cell
in Fig. \ref{fig:(a)-Geometry-and}(a). Note that the helices in this
design have no electrical connection. A periodic array composed of
an infinite number of such chiral unit cells with $d=\sqrt{2}\lambda/5$
where $\lambda$ is the plane wave's wavelength shows a perfect $y$
to $x$ polarization rotation at the frequency of $1.5\,\mathrm{GHz},$as
shown in Fig. \ref{fig:(a)-Geometry-and}(d), since the reflection
and transmission coefficients of such a metasurface satisfy Eqs. (\ref{eq:refless_MS}).
Deduced from this figure, the incident wave with its electric field
polarization along $y$ is perfectly transmitted into a transmitted
wave with electric field polarization along $x$, propagating in the
lower half space. It is obvious from this Figure that at the $1.5\,\mathrm{GHz},$
both the $x$- and $y$-polarized reflected waves are negligible.

\subsection{Supercell design for wavefront deflection}

In this section we use a modulation of the metasurface parameters
that based on the generalized Snell's law leads to the transmission
phase in Eq. (\ref{trans_phase}). Following this approach, the shape
of the wavefront refracted by the array relies on the gradual increase
of the transmission phase along the supercell constitutive elements
\cite{Yu2011,veysi_thin_2015,khorasaninejad2015achromatic,estakhri2016recent,Wong2016}.
Such gradual phase increase along the metasurface is provided by suitably
engineering each constitutive inclusion in the so called ``supercell'',
and since the phase variation is along the \textit{$y$} direction,
the supercell is defined by modulating a few unit cells along the
\textit{$y$} direction, while it has the dimension of a single unit
cell in the \textit{$x$} direction. According to the example in the
previous section and Eq. (\ref{trans_phase}), considering a normal
incidence, a required $45^{\circ}$ transmission angle deflection,
and assuming $\varphi_{t}=0$ without loss of generality, the transmission
phase shall increase linearly in the \textit{$y$} direction as 

\begin{equation}
\Phi_{t}(y)=-\frac{k_{0}}{\sqrt{2}}y.\label{T_phase_change}
\end{equation}

This required transmission phase is a continuous function of position
$y$ along the surface. However, as discussed before, for a practical
realization the continuous distribution of electric and magnetic polarization
densities is realized in a discretized fashion, i.e., by a finite
number of electric and magnetic induced dipole moments in each supercell.
The desired phase distribution across the metasurface is obtained
by meticulously optimizing the dimensions of the four interlaced helices
in a few unit cells that make the supercell. In the present realization
the supercell is divided into five unit cells as shown in the figure
below Table \ref{table_of_inclus_params}, that shows the optimized
dimensions of the helices in each of the five unit cells. The required
transmission phases at the location of the five different unit cells
are obtained from Eq. (\ref{T_phase_change}) where $y=\pm nd$ where
$n=0,1,2$. Note that all the four helices in each individual unit
cell are identical. 

\begin{figure}[t]
\begin{centering}
\includegraphics[scale=0.6]{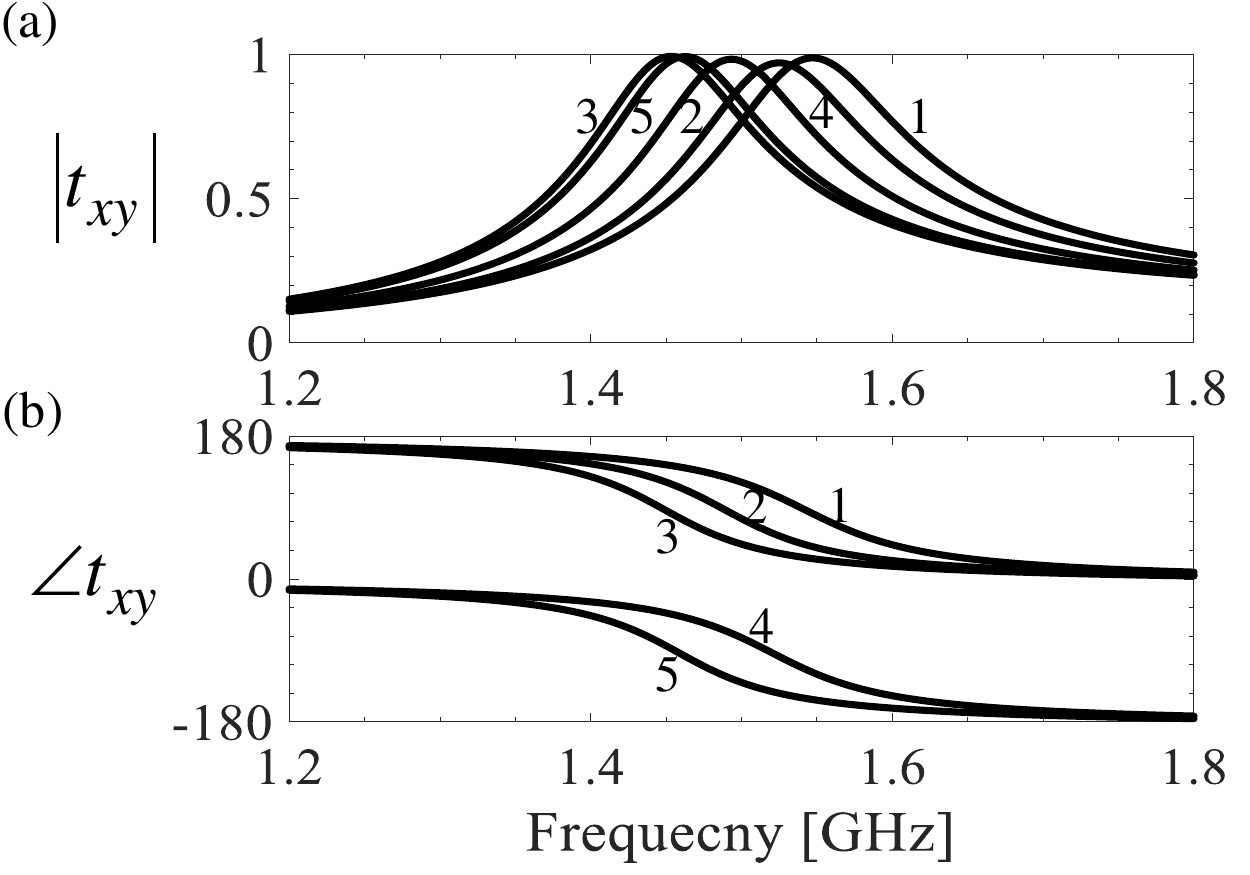}
\par\end{centering}
\caption{Magnitude and phase of $t_{xy}$ transmission coefficient for the
five different metasurfaces, each one based on one unit cell design
in Table \ref{table_of_inclus_params}: (a) Magnitude and (b) Phase
of the cross polarization transmission coefficient $t_{xy}$ . When
it reaches 1 all the power is transmitted with a polarization rotation
of $90^{\circ}$ as discussed in Sec. \ref{subsec:Unit-cell-design}.
The phase gradient metasurface is then made with a supercell with
dimension $d\times5d$ , by arranging sequentially the five properly
designed unit cells.}
\label{super_cell_Tcr}
\end{figure}

\begin{table}[t]
\begin{centering}
\begin{tabular}{c|c|c|c|c|c}
 & \# 1 & \# 2 & \# 3 & \# 4 & \# 5\tabularnewline
\hline 
\hline 
$r\,(\mathrm{mm})$  & $13.23$ & $13.76$ & $14.24$ & $13.30$ & $14.24$\tabularnewline
\hline 
$p\,(\mathrm{mm})$ & $17.13$ & $17.8$ & $18.43$ & $17.22$ & $18.43$\tabularnewline
\hline 
$handedness$ & $L$ & $L$ & $L$ & $R$ & $R$\tabularnewline
\hline 
$\Phi_{t}\,(\mathrm{deg.})$ & $120$ & $80$ & $41$ & $-73$ & $-140$\tabularnewline
\end{tabular}
\par\end{centering}
\begin{centering}
\includegraphics[width=3.5in]{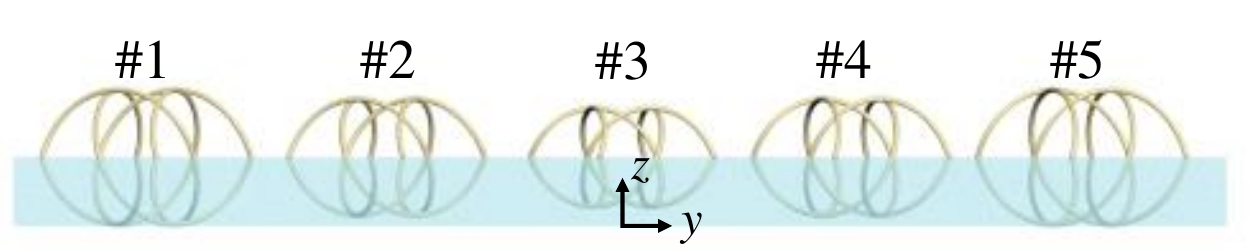}
\par\end{centering}
\centering{}\caption{Helix radius, helix pitch and the handedness of the 5 different inclusions
(helices are identical in each cluster). In the table $R,L$ denote
the right and left handed helices. }
\label{table_of_inclus_params}
\end{table}
Therefore the metasurface is made of an array of supercells with dimension
$d$ along the $x$ direction and $5d$ along the $y$ direction.
The design of the supercell elements is done as follows: First, we
design five distinct metasurfaces, each one made of a periodic array
of each unit cells in table \ref{table_of_inclus_params}, with period
$d$ in both $x$ and $y$ directions. The full-wave simulation based
on the finite element method is based on periodic boundary conditions.
Therefore, a single unit cell of dimension $d\times d$ is simulated
for normal plane wave incidence, and the transmitted phase is evaluated
for each type of these five metasurfaces. For each metasurface, dimensions
are optimized to provide a perfect cross-polarized transmittance (this
is possible based on the results of the previous subsection) and transmission
phase given in Table \ref{table_of_inclus_params}. The resulting
magnitude and phase of the $y$- to $x$-polarized wave transmission
coefficient $t_{xy}$ is depicted in Fig. \ref{super_cell_Tcr} for
the five different metasurfaces, each one based on a different inclusion's
dimensions given in Table \ref{table_of_inclus_params}. 

\begin{figure}[t]
\begin{raggedright}
(a)
\par\end{raggedright}
\begin{centering}
\includegraphics[scale=0.9]{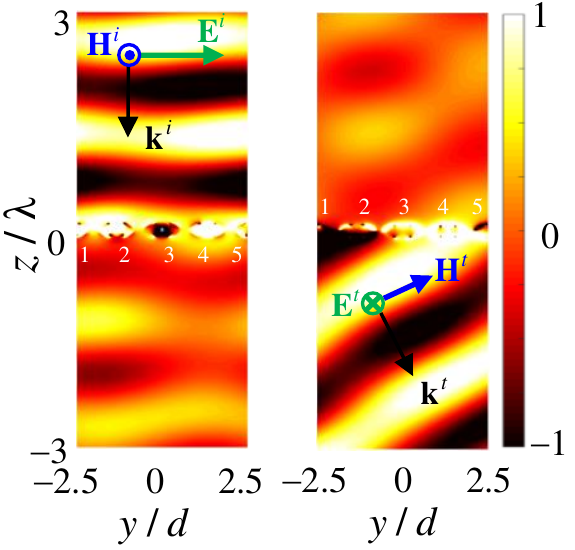} 
\par\end{centering}
\begin{raggedright}
(b)
\par\end{raggedright}
\begin{centering}
\includegraphics[scale=1.4]{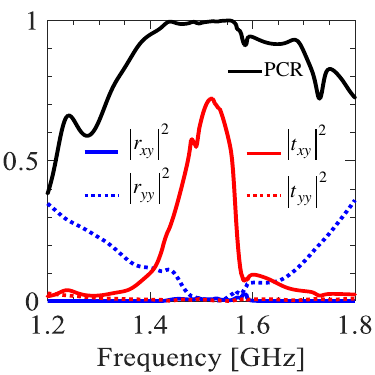}
\par\end{centering}
\caption{(a) Full wave simulation for the field distribution resulting from
normal plane wave incidence from the top on the designed metasurface
whose supercell is made of five unit cells with parameters given in
Table \ref{table_of_inclus_params}. (Left) \textit{\textcolor{black}{y}}-polarization
of the electric field showing mainly the incident. (Right) \textit{\textcolor{black}{x}}-polarization
of the electric field showing mainly the deflected transmitted field.
It is clear that the metasurface besides deflecting the wavefront
also rotates the polarization by $90^{\circ}$ degrees. (b) Plot of
the PCR, the $y$- to $x$-polarized power reflection and power transmission
coefficients versus frequency showing an almost $100\%$ polarization
rotation, and the $y$- to $x$-polarized power transmission coefficient
(i.e, the the cross polarized power transmission) showing a $72\%$
power transmission efficiency. Metal and dielectric losses are accounted
for in this simulation.}
\label{CST_fields}
\end{figure}

Next, the gradient metasurface to provide an angular deflection of
$45^{\circ}$ is made of a periodic arrangement of supercells with
dimension $d$ along the $x$ direction and $5d$ along the $y$ direction.
As a result the designed metasurface has a desired periodic phase
distribution along the $y$ direction (see Eq. (\ref{T_phase_change}))
and a uniform phase distribution along the $x$ direction. The field
distribution in proximity of the engineered gradient metasurface,
excited with a normally incident plane wave polarized along the $y$\textcolor{blue}{{}
}direction, is illustrated in Fig. \ref{CST_fields}(a) where the
field distribution is calculated with a full wave simulation based
on the finite element method. \textcolor{black}{The} distribution
of the $y$-polarization (left) and the $x$-polarization (right)
of the electric field are plotted in Fig. \ref{CST_fields}(a) at
the frequency of $f=1.5\,\mathrm{GHz}$. In this simulation based
on the finite\textcolor{black}{{} element method, periodic boundaries
are chosen to mimic the infinite extension} of the metasurface along
both $x$ and $y$ directions, i.e., modeling a periodic supercell
of dimensions $d\times5d$.\textcolor{blue}{{} }In this combined design
we account for losses, i.e., helices are made of copper and are embedded
in a foam (FR 3703 from General Plastics) with dielectric constant
of 1.06 and loss tangent of 0.0004 as in Fig. \ref{fig:(a)-Geometry-and}.
As it is clear from Fig. \ref{CST_fields}, the gradient metasurface
rotates the polarization of the incident wave by $90^{\circ}$ and
refracts it into $\theta_{t}=45^{\circ}$.\textcolor{red}{{} }The result
in Fig. \ref{CST_fields}(a) does not show the perfect polarization
rotation and deflection as the ideal case in Fig. \ref{fig:(a)-Geometry,-incident,}
because in the actual design of the gradient metasurface, we have
used the concept of local periodicity in designing the five unit cells
of the supercell, which is a standard approximation in metasurface
and reflectarray design \cite{huang_reflectarray_2007,veysi_thin_2015}
but it is not fully accurate. As a measure of the metasurface performance
we define the polarization conversion ratio (PCR) in transmission
as

\begin{equation}
\mathrm{PCR}=\frac{\left|t_{xy}\right|^{2}}{\left|t_{yy}\right|^{2}+\left|t_{xy}\right|^{2}}
\end{equation}

\noindent where $\left|t_{xy}\right|$ and $\left|t_{yy}\right|$
are the magnitude of $y$- to $x$-polarized (cross-pol component)
and $y$- to $y$-polarized (co-pol component) transmission coefficients,
respectively. Fig. \ref{CST_fields}(b) shows the PCR versus frequency,
as well as the $y$- to $x$-polarized reflectance and transmittance
of a $y$-polarized incident wave. As it is obvious from this figure
we obtain a perfect ($100\%$) polarization conversion and $72\%$
power transmission into the deflected wave with the engineered metasurface.
To the best of our knowledge, this is an unprecedented result, i.e.,
this is the first design of a metasurface that provides simultaneous
deflection and polarization rotation of the incoming wavefront with
such a high efficiency.

\section{Conclusion}

In the framework of gradient metasurfaces we have shown that in principle,
using chirality, it is possible to obtain perfect polarization rotation
of the electromagnetic wavefront with concurrent full transmission
into a desired deflected direction. The chirality characteristic of
a metasurface serves for polarization rotation of the transmitted
plane wave with respect to the incident one, whereas the gradient
property (i.e., the spatial dispersion) of the metasurface grants
for the wavefront deflection. Furthermore, we have demonstrated a
possible physical realization of the proposed device by engineering
a proper metasurface unit cell inclusion that realizes the aforementioned
combined functionalities. Our full wave simulation results demonstrate
high transmission power efficiency of $72\%$ at an angle of $45^{\circ}$
by using only one single layer of inclusions (i.e., a single metasurface)
which is accompanied by a perfect $90^{\circ}$ polarization rotation.
Despite the fact that the results are shown for a specific illustrative
case, the method outlined in this paper is very general and can be
used for conceiving metasurfaces that deflect wavefronts at any angle
with arbitrary polarization conversion and with (in theory) perfect
transmittance. 

In short, in a single metasurface we have combined two interesting
functionalities, i.e., a wave refraction at a given angle and a polarization
rotation, with very high efficiency and subwavelength thickness, by
using chiral metasurface inclusions.

\section{Acknowledgment}

The authors would like to thank DS SIMULIA for providing CST Studio
Suite that was instrumental in this study. Also, the authors acknowledge
support from the W. M. Keck Foundation, USA.

\appendix

\section{Energy densities of the incident and refracted waves, power balance
relation, and transmission coefficient for perfect power transmission
and polarization conversion\label{AppB}}

\begin{figure}[t]
\begin{centering}
\includegraphics[width=2.3in]{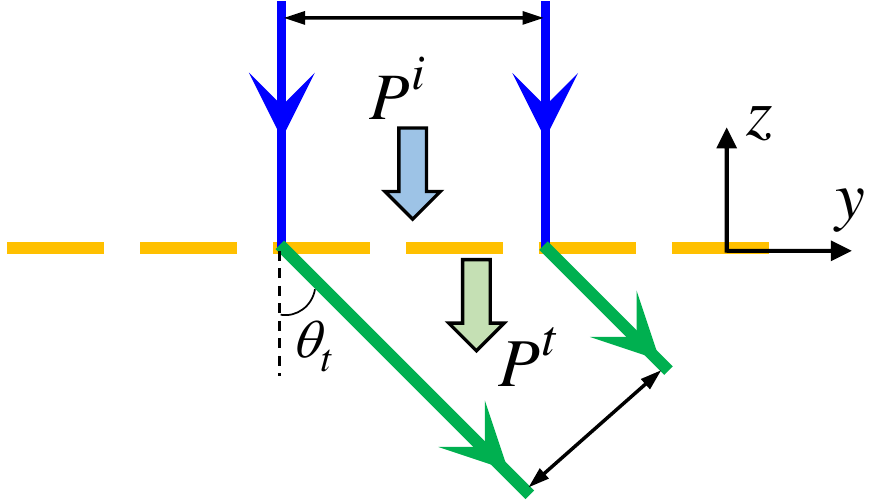}
\par\end{centering}
\caption{\label{fig:Schematic-of-the}Schematic of the power flow for a propagating
wave when passing through the deflecting metasurface. The metasurface
is passive and changes the polarization and the direction of the transmitted
wavefront with respect to the incident one.}
\end{figure}

The characteristic of the proposed fully-transmissive metasurface
is such that the wave passes through it without changing its total
time-average power (when the metasurface is lossless) and the metasurface
changes the polarization and the direction of the transmitted wavefront
with respect to the incident one. Therefore, the total incident and
refracted powers crossing the metasurface along the ray tube in Fig.
\ref{fig:Schematic-of-the} are equal. Under such condition, the\textcolor{blue}{{}
}total time-average incident and transmitted field energy densities
are required to satisfy 

\begin{equation}
\frac{\mathscr{\mathcal{E}}^{i}}{\mathscr{\mathcal{E}}^{t}}=\frac{\cos\theta_{t}}{\cos\theta_{i}}.
\end{equation}

Therefore in this scenario, since the total time-average incident
and transmitted energy densities, $\mathscr{\mathcal{E}}^{i}$ and
$\mathscr{\mathcal{E}}^{t}$, are different, the wave intensity changes
when passing through the fully-transmissive metasurface. This can
be described also in the following manner, by observing the power
crossing a plane with constant $z$, right above and below the metsurface.
Of course such powers are equal to each other when considering a lossless
metasurface, which implies the conservation of the normal component
of the Poynting vector from above to below the metasurface, i.e.,

\begin{equation}
\frac{1}{2}\mathrm{Re}\left(\mathbf{E}_{t+}\times\mathbf{H}_{t+}^{\ast}\right)=\frac{1}{2}\mathbb{\mathrm{Re}}\left(\mathbf{E}_{t-}\times\mathbf{H}_{t-}^{\ast}\right).\label{eq:power balance-1}
\end{equation}

Substituting fields described in (\ref{E_t}) and (\ref{H_t}) into
(\ref{eq:power balance-1}) leads to

\begin{equation}
\frac{E_{y}^{i\,2}}{\cos\theta_{i}}=\cos\theta_{t}E_{x}^{t\,2},\label{eq:pwr balance 2-1}
\end{equation}
where $E_{y}^{i}=E_{0}^{i}\cos\theta_{i}$ and $E_{x}^{t}=E_{0}^{t}$,
then

\begin{equation}
t_{xy}=\frac{1}{\sqrt{\cos\theta_{i}\cos\theta_{t}}},
\end{equation}
for a reflectionless surface. Thus for normal incidence, $t_{xy}=1/\sqrt{\cos\theta_{t}}$
which is larger than unity for an oblique transmission angle, without
contradicting the power conservation law.

\bibliographystyle{apsrev4-1}
\bibliography{Chiral_MS_v3}

\end{document}